\documentclass[pra,twocolumn,superscriptaddress,a4paper,english,longbibliography]{revtex4-2}

\usepackage{times}
\usepackage{color}
\usepackage[colorlinks=true,urlcolor=blue,citecolor=blue]{hyperref}
\usepackage{url}
\usepackage{breakurl}
\usepackage{soul}
\usepackage{graphicx}
\usepackage{amsmath}
\usepackage{subfigure}
\usepackage{xcolor}
\usepackage{amssymb}
\usepackage{hyperref}
\DeclareGraphicsExtensions{.png,.eps}
\usepackage{float}
\usepackage{appendix}
\usepackage{etoolbox}
\usepackage{algorithm}  
\usepackage{algpseudocode}  

\newcommand {\R}{\textcolor {black}}

\usepackage{xcolor}
\usepackage[urlcolor=blue]{hyperref}      
\hypersetup{
    colorlinks = true,                    
    citecolor = {blue},
    linkcolor = {purple},
           }

\begin{document}	
	
\title{Preparation of Many-body Ground States by Time Evolution with Variational Microscopic Magnetic Fields and Incomplete Interactions}

\author{Ying Lu}
\affiliation{Department of Physics, Capital Normal University, Beijing 100048, China}
\author{Yue-Min Li}
\affiliation{Department of Physics, Capital Normal University, Beijing 100048, China}
\author{Peng-Fei Zhou}
\affiliation{Department of Physics, Capital Normal University, Beijing 100048, China}
\author{Shi-Ju Ran} \email[Corresponding author. Email: ] {sjran@cnu.edu.cn}
\affiliation{Department of Physics, Capital Normal University, Beijing 100048, China}
\date{\today}

\begin{abstract}
	State preparation is of fundamental importance in quantum physics, which can be realized by constructing the quantum circuit as a unitary that transforms the initial state to the target, or implementing a quantum control protocol to evolve to the target state with a designed Hamiltonian. In this article, we study the latter on quantum many-body systems by the time evolution with fixed couplings and variational magnetic fields. Specifically, we consider preparing the ground states of the Hamiltonians containing certain interactions that are missing in the Hamiltonians for the time evolution. An optimization method is proposed to optimize the magnetic fields by ``fine-graining'' the discretization of time, in order to gain high precision and stability. The \R{automatic differentiation} technique is utilized to obtain the gradients of the fields against the logarithmic fidelity. Our method is tested on preparing the ground state of the Heisenberg chain with the time evolution by the XY and Ising interactions, and its performance surpasses two baseline methods that use local and global optimization strategies, respectively. Our work can be applied and generalized to other quantum models such as those defined on higher dimensional lattices. It enlightens to reduce the complexity of the required interactions for implementing quantum control or other tasks in quantum information and computation by means of optimizing the magnetic fields. 

\end{abstract}

\maketitle

\section{Introduction}

How to efficiently and accurately obtain the desired states on quantum systems belongs to the fundamental topics in the fields of condensed matter physics, quantum simulation, quantum computation, and beyond. Taking the strongly correlated quantum systems as an example, a class of states with non-trivial properties, such as quantum spin liquids with possible topological orders~\cite{M00QSLrev2000, B10QSLRev2010, W12QSL2012,SB17QSLRev2017}, can be reached by finding in nature or synthesizing the materials with the expected interactions, such as the antiferromagnets on two-dimensional lattices with geometrical frustration~\cite{MR06FrustrateRev2006}. Usually, the target state is the ground or low-lying excited state of the Hamiltonian, which thus can be reached by annealing~\cite{PREQAITIMKTN1998}.

With controllable parameters in the Hamiltonian, the state of a quantum system can be driven to a specific target by evolution (see an early work in Ref.~\cite{JMPOCQMSHGM1983} and a review in Ref.~\cite{Control2010rev} as examples). For instance, a molecule can be driven to the desired state by designing a sequence of laser pulses optimized according to the fitness values from certain measurements on the state of the molecule~\cite{CPQMOCPBIM1990,PRLQCSPJPR1992}. The optimization of the parameters of the time evolution can be formulated as optimal control problems, which have been widely studied in, e.g., interacting spins and solids~\cite{QCent2014nc, CrtSolid2016, PRLGPOQCDAG2019, CtrSolid2020}. Typical approaches for quantum control include solving the quantum brachistochrone equations to obtain the evolution to a target state with minimal time cost~\cite{PRLTOQECAHAKTOY2006, Brachistochrone2015}.

In recent years, machine learning (ML) has shed new light on developing efficient protocols for state preparation and quantum control. One popular trend is to use reinforcement learning, aiming for short, high-fidelity driving protocols or something similar \cite{PRXRLDPQCBUKOV2018, PRBRLAPFESIQKOBM2018, PRAASCLETQSLZXM2018, niu2018universal, RL2020PRL, RL2020PRR, RL2020QMI, RL2020PRALXP, RLstate2021PRL}. It has been proposed to adapt the ML models, such as deep neural networks, to generate or optimize the controlling parameters~\cite{PRANMDPSRSTQyangxu2018, LRQC2019PRA, ostaszewski2019QIP, QCAG2020IOP}. In particular, \R{automatic differentiation}, a widely applied numerical technique, has been applied to construct large-scale quantum circuits for state preparation~\cite{ADQC2021ZPF}, speed up the numerical simulations of optimal quantum control~\cite{BPQC2017PRA}, and to other topics such as variational quantum eigen-solver~\cite{ADVQE2020PRB} and tensor network simulations~\cite{PRXDPTNwanglei2019,dTRG2020}.

\begin{figure}[tbp]
	\centering
	\includegraphics[angle=0,width=1\linewidth]{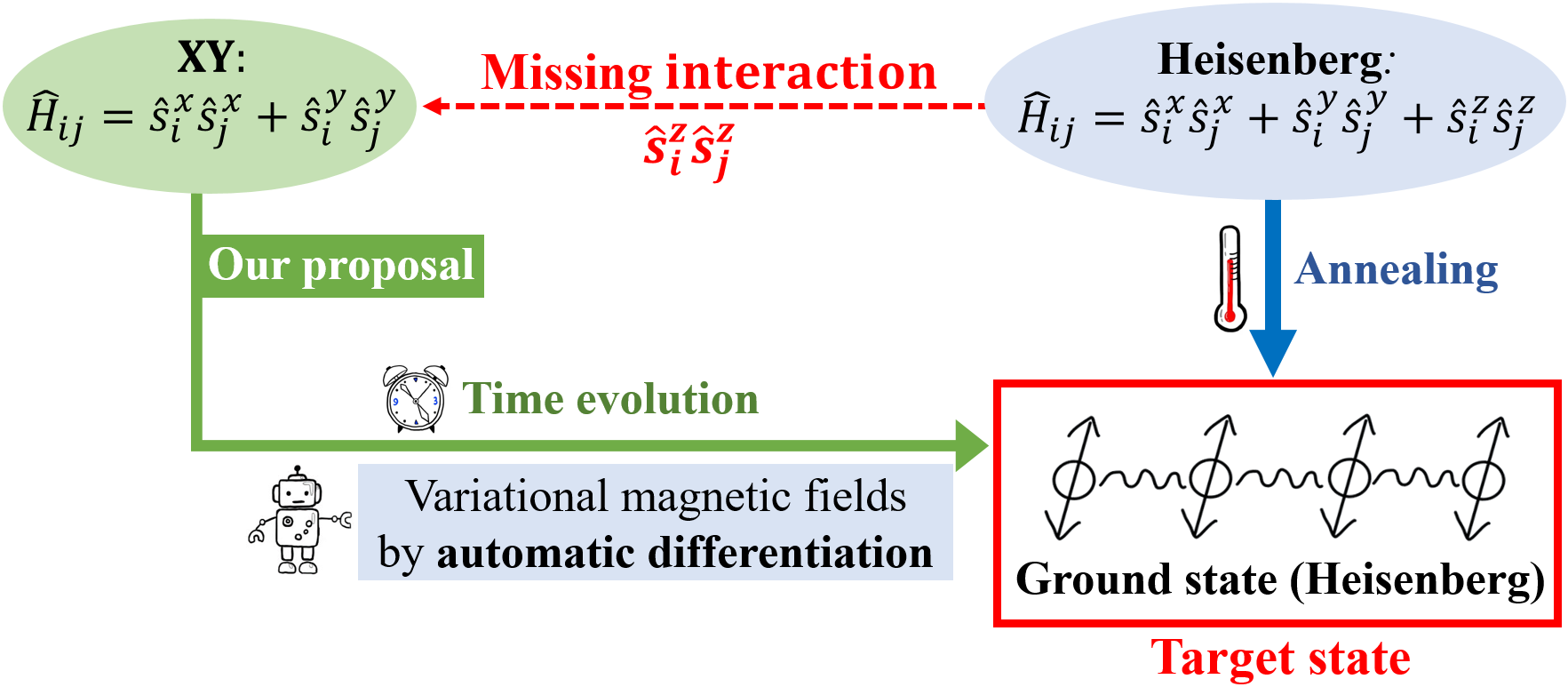}
	\caption{\R{The illustration of state preparation by time evolution with variational magnetic fields. Here, we take the target state as the ground state of the Heisenberg chain as an example. For a given Hamiltonian, the ground state can be obtained in an annealing process. We propose to tune the magnetic fields to evolve from a product state to the target state, while the Hamiltonian for time evolution contains less interaction terms (such as the XY or Ising interactions). The magnetic fields are optimized using the automatic differentiation technique that was originally developed in machine learning.}}
	\label{fig-idea}
\end{figure}

\begin{figure}[tbp]
	\centering
	\includegraphics[angle=0,width=1\linewidth]{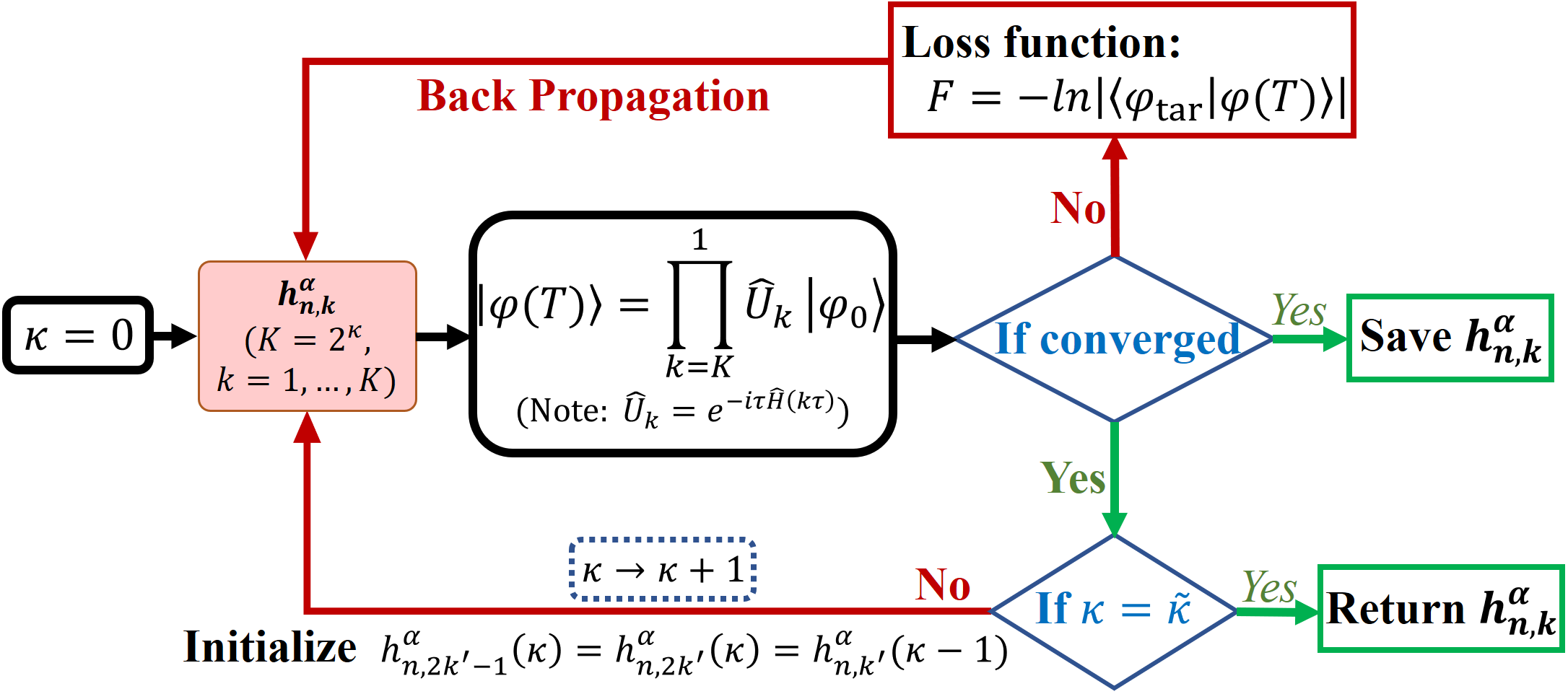}
	\caption{The flowchart of fine-grained time optimization. \R{The letter $K$ represents the total number of the time slices in a time evolution. We have $T=K\tau$, with $T$ the total evolution time and $\tau$ the time interval for each slice. The letter $\tilde{\kappa}$ represents the total number of fine-grained steps. For the $\kappa$-th fine-graining step with $\kappa = 0, \ldots, \tilde{\kappa}$, we have $K=2^\kappa$. The letter $k$ labels the time slice in a specific time evolution and takes $k=1, \ldots, K$. The iteration shown in the upper half of the figure is the optimization of magnetic fields for a fixed $K$. The iteration shown by the lower half is to gradually fine grain time by increasing $K$ to $2K$. }}
	\label{fig-flowchart}
\end{figure}

In this article, we consider the preparation of the ground states of quantum many-body systems by time evolution with fixed but incomplete interactions and variational magnetic fields. The idea is illustrated in Fig. \ref{fig-idea}. We assume that the Hamiltonian for the time evolution does not contain all the interaction terms in the Hamiltonian that gives the ground state as our target. The fine-grained time optimization (FGTO) algorithm is proposed by ``fine-graining'' the discretization of time, i.e., by gradually increasing the allowed maximal frequency of the time-dependent magnetic fields. The negative logarithmic fidelity is minimized by optimizing the fields with a gradient descent. We utilize the \R{automatic differentiation} technique to obtain the gradients of the fields. The flowchart to illustrate the FGTO is given in Fig. \ref{fig-flowchart}.

To benchmark FGTO, we consider preparing the ground state of the Heisenberg model by time evolution with XY or Ising interactions in space- and time-dependent magnetic fields. Note the Heisenberg model contains two-body interactions along three spin directions, while the XY and Ising models contain only interactions in one or two spin directions. The FGTO is compared with two baseline methods where local and brutal-force global optimizations of the magnetic fields are implemented, respectively. FGTO achieves the highest preparation precision among these three methods. Numerical results are provided to show how the precision varies with the total evolution time and the fineness of the time discretization.

\section{General scheme of time-evolution state preparation}

We consider the preparation of the state $|\psi_{\text{tar}}\rangle$ by time evolution of the Hamiltonian
\begin{eqnarray}
\hat{H}(t) = \sum_{m,n} \hat{H}_{mn} + \sum_n \sum_{\alpha=x,y,z} h^{\alpha}_n(t) \hat{S}_n^{\alpha}.
\label{eq-H}
\end{eqnarray}
with $\hat{S}_n^{\alpha}$ the spin operator on the $n$-th site and $h^{\alpha}_n(t)$ the magnetic fields at time $t$. We take the two-body interaction terms in the form of $\hat{H}_{mn} = \sum_{\alpha=x,y,z} J^{\alpha} \hat{S}^{\alpha}_m \hat{S}^{\alpha}_n$. Here, the Hamiltonians are assumed to contain only nearest-neighbor couplings and the coupling constants ($J^{\alpha}$) are assumed to be space- and time independent.

Our goal is to obtain optimal magnetic fields $h^{\alpha}_n(t)$ that minimize the distance between the target state $|\psi_{\text{tar}} \rangle$ and the evolved state
\begin{eqnarray}
|\psi(T)\rangle = \hat{U}(T) |\psi_0\rangle,
\label{eq-psit}
\end{eqnarray}
with $\hat{U}(T)$ formally denoting the time evolution operator determined by $\hat{H}(t)$, $|\psi_0\rangle=|\psi(t=0)\rangle$ the initial state, and $T$ the total evolution time. The distance can be characterized by negative logarithmic fidelity (NLF) per site as
\begin{eqnarray}
\R{F = - \ln f,}
\label{eq-F}
\end{eqnarray}
\R{where $f$ is the fidelity defined as}
\begin{eqnarray}
	\R{f=\left|\langle \psi_{\text{tar}}|\psi(T) \rangle \right|,}
	\label{eq-f}
\end{eqnarray}
We require knowledge on the evolved state to evaluate $F$ in the optimization, which is in a similar case of adiabatic tracking~\cite{AdiaTrack2014PRA}. Essentially, our controlling scheme can be generalized from fidelity to observables, so that the full wave function will not be needed to optimize the evolution Hamiltonian.

In practice, we discretize the total time $T$ to $K$ identical slices, and the evolved state can be approximated as
\begin{eqnarray}
|\psi(T)\rangle &=& e^{-i\tau \hat{H} (K\tau)} \ldots  e^{-i\tau \hat{H} (2\tau)} e^{-i\tau \hat{H} (\tau)} |\psi_0\rangle  \nonumber  \\
&=& \prod_{k=K}^{1} e^{-i\tau \hat{H} (k\tau)} |\psi_0\rangle.
\label{eq-psit1}
\end{eqnarray}
with $\tau = \frac{T}{K}$. It means during the time of $(k-1)\tau \leq t < k\tau$, we assume that $h^{\alpha}_n(t)$ does not change and takes the value $h^{\alpha}_n(t) = h^{\alpha}_{n,k}$, which belongs to the generalizations of the bang-bang protocols \cite{BangBang1, BangBang2}. Without losing generality, we take the initial state $|\psi_{0}\rangle = \prod_{\otimes n=1}^N |0_n\rangle$ with $|0_n\rangle$ the spin-up state of the $n$-th spin.

\section{Fine-grained time optimization}

We use the gradient descent to update the magnetic fields as
\begin{eqnarray}
h^{\alpha}_{n,k} \leftarrow h^{\alpha}_{n,k} - \eta \frac{\partial F}{\partial h^{\alpha}_{n,k}},
\label{eq-updateh}
\end{eqnarray}
\R{where the gradients $\frac{\partial F}{\partial h^{\alpha}_{n,k}}$ are obtained by the automatic differentiation of PYTORCH \cite{PyTorch}, and $\eta$ represents} the learning rate controlled by the Adam optimizer \cite{KB15Adam}. $F$ is also called the loss function. However, our simulations show that this optimization problem possesses many local minima. Consequently, the results might be sensitive to the initial values of $h^{\alpha}_{n,k}$. Therefore, the initialization strategy becomes crucial.

We propose the fine-grained time optimization (FGTO) algorithm (Fig. \ref{fig-flowchart}). To begin with, we set $K=1$ in the first iteration $\kappa=0$, meaning we do not allow $h^{\alpha}_{n,k}(\kappa=0)$ (note $k=1$ and $n=1, \cdots, N$) to change in the whole evolution. \R{The initial magnetic fields are taken randomly.} The optimal values of $h^{\alpha}_{n,k}(\kappa=0)$ are reached by implementing the gradient optimization using Eq. (\ref{eq-updateh}) for sufficiently many times.

After $h^{\alpha}_{n,k}(\kappa=0)$ converges, we ``fine-grain'' the time discretization by increasing $K$ to $2K$ for the iteration of $\kappa=1$. The magnetic fields are initialized as $h^{\alpha}_{n,k}(\kappa=1) = h^{\alpha}_{n,1}(\kappa=0)$ for $k=1$ and $2$. For the iteration of $\kappa>1$, we initialize the magnetic fields according to those obtained in the ($\kappa-1$)-th iteration as $h^{\alpha}_{n,2k'-1}(\kappa) = h^{\alpha}_{n,2k'}(\kappa) = h^{\alpha}_{n,k'}(\kappa-1) $, which are subsequently updated by the gradient optimization.

To compare with, we also try another two optimization algorithms as the baselines. The first is dubbed the sliced time optimization (STO), where $h^{\alpha}_{n,k}$ are optimized slice by slice from $k=1$ to $K$. The idea is to minimize the loss function for every time slice, \R{in the same spirit of the ``greedy'' algorithms}. In the optimization of the $k$-th slice, $h^{\alpha}_{n,k'}$ for $k'<k$ are fixed as the values obtained in the former iterations. The fields in the $k$-th time slice $h^{\alpha}_{n,k}$ are optimized by minimizing the distance between the target state and the evolved state at $t = k\tau$. The loss function \R{is taken as $F_k =  - \ln \left|\langle \psi_{\text{tar}}|\psi(k\tau) \rangle \right|$}. 

The other method is called as global time optimization (GTO), where we just iteratively compute the loss of $|\psi(T)\rangle$ in Eq. (\ref{eq-F}), and update $h^{\alpha}_{n,k}$ for all $k=1$ to $K$ simultaneously. Compared with GTO, STO is a more economical method since in each iteration, we only need to deal with the computational graphs of the automatic differentiation for the $h^{\alpha}_{n,k}$ in a single time slice. \R{Since STO optimizes the magnetic fields by locally considering the evolution in one time slice, it can be trapped to some local minimum. Consequently, GTO shows higher accuracy than STO.}  For FGTO, one can see that we in fact use GTO to update $h^{\alpha}_{n,k}$ with a smaller or (eventually) equal number of time slices. \R{Furthermore in FGTO, the fields are properly initialized in the fine-grain process of time. For GTO in comparison, we need to guess $3K$ values to initialize the fields. Even though we use STO to initialize $h^{\alpha}_{n,k}$, the results are still worse than those by FGTO, possibly because STO already drive the optimization into a local minimum. In all, FGTO better balances the efficiency and accuracy, and is more stable by using a more reasonable strategy in initialization. See the pseudo codes and flowcharts of these algorithms in the Supplemental Material \cite{sm}.}

Compared with the quantum control schemes with reinforcement learning (such as Refs. [\onlinecite{PRXRLDPQCBUKOV2018, PRBRLAPFESIQKOBM2018, niu2018universal, RL2020PRL, RL2020PRR, RL2020QMI, RL2020PRALXP, RLstate2021PRL}]), one advantage of our scheme is its simplicity. As we directly optimize the time-dependent magnetic fields by automatic differentiation, we are not concerned that the results might be affected by the choices of neural network (NN) with different learning and generalization abilities. Meanwhile, training and testing samples are not needed. An advantage of the reinforcement-learning control schemes is that once the NN is well trained, it can be used to solve a class of optimal control problems. The range of problems that a trained NN can reliably solve (without re-training by new samples) would depend on its generalization power. Therefore, our proposal and the reinforcement-learning schemes can complement each other to solve specific problems in the field of quantum control.

\section{Numerical results}

To show the validity of state preparation by time evolution, we choose the target state $|\psi_{\text{tar}}\rangle$ as the ground state of the one-dimensional (1D) Heisenberg model (HM) with the Hamiltonian
\begin{eqnarray}
	\hat{H}_{\text{HM}} = \sum_{n=1}^{N-1} \sum_{\alpha=x,y,z} \hat{S}_n^{\alpha} \hat{S}_{n+1}^{\alpha},
	\label{eq-Heisenberg}
\end{eqnarray}
where we take the number of spins $N=10$ as an example. Its ground state is a spin liquid with no magnetic ordering. For the Hamiltonian to implement time evolution, we choose the XY model and quantum Ising model (QIM) as examples, which read
\begin{eqnarray}
	\hat{H}_{\text{XY}}(t) &=& \sum_{n=1}^{N-1} \sum_{\alpha=x,y} \hat{S}_n^{\alpha} \hat{S}_{n+1}^{\alpha} + \sum_{n=1}^N \sum_{\alpha=x,y,z} h^{\alpha}_n(t) \hat{S}_n^{\alpha},  \\
	\hat{H}_{\text{QIM}}(t) &=& \sum_{n=1}^{N-1}  \hat{S}_n^{z} \hat{S}_{n+1}^{z} + \sum_{n=1}^N \sum_{\alpha=x,y,z} h^{\alpha}_n(t) \hat{S}_n^{\alpha}.
	\label{eq-Hevolve}
\end{eqnarray}
Our aim is to see by tuning the magnetic fields, how accurately the time evolution by $\hat{H}_{\text{XY}}(t)$ or $\hat{H}_{\text{QIM}}(t)$ can prepare the ground state of $\hat{H}_{\text{HM}}$, with the fact that some coupling terms in $\hat{H}_{\text{HM}}$ are missing in $\hat{H}_{\text{XY}}(t)$ or $\hat{H}_{\text{QIM}}(t)$.

\begin{figure}[tbp]
	\centering
	\includegraphics[angle=0,width=0.8\linewidth]{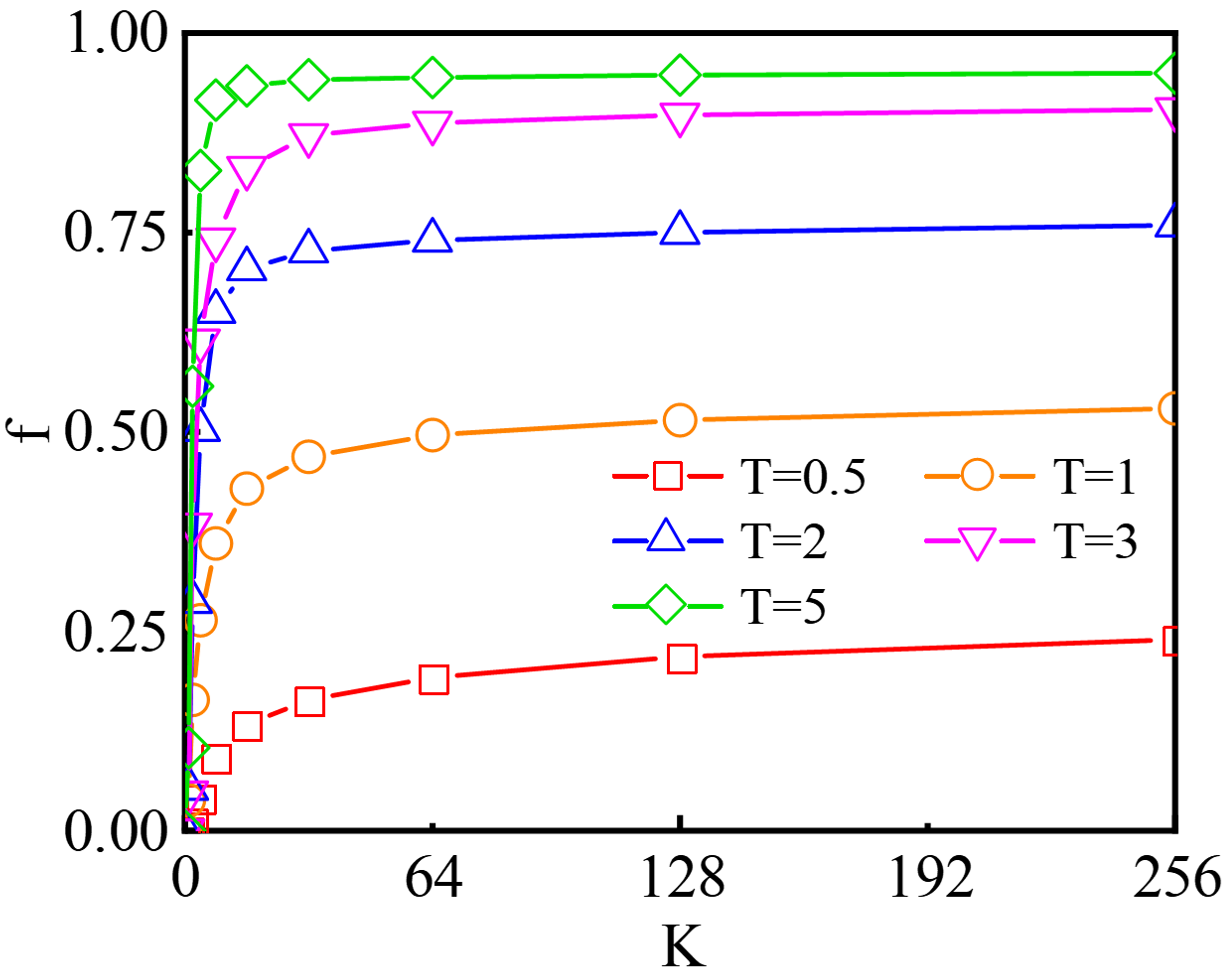}
	\caption{(Color online) The \R{fidelity $f$ in Eq. (\ref{eq-f})} vs the number of time slices $K$. The target state is the ground state of the Heisenberg model and the evolution Hamiltonian is taken as the XY model. We employ the FGTO method, and \R{take the number of spins $N=10$ and} the total evolution time $T=0.5$, $1$, $2$, $3$, and $5$.}
	\label{fig-FvsK}
\end{figure}

Figure \ref{fig-FvsK} demonstrates \R{$f$ [Eq. (\ref{eq-f})]} by increasing the number of time slices $K$. For $K=1$ \R{(i.e., $T=\tau$)}, meaning the magnetic fields are not allowed to vary with time, \R{we have a small $f$ (with $F \simeq 3$)}. By increasing $K$, we allow the magnetic fields to change more frequently. \R{The fidelity $f$ increases quickly} with $K$ and approximately converges at about $K=32$. Note the target state $|\psi_{\text{tar}}\rangle$ and the initial state $|\psi(0)\rangle$ are almost orthogonal to each other with the NLF $F \simeq 60$. This is partially because the fidelity usually decreases exponentially with the number of qubits $N$ and we take $N \gg 1$.

\begin{figure}[tbp]
	\centering
	\includegraphics[angle=0,width=0.8\linewidth]{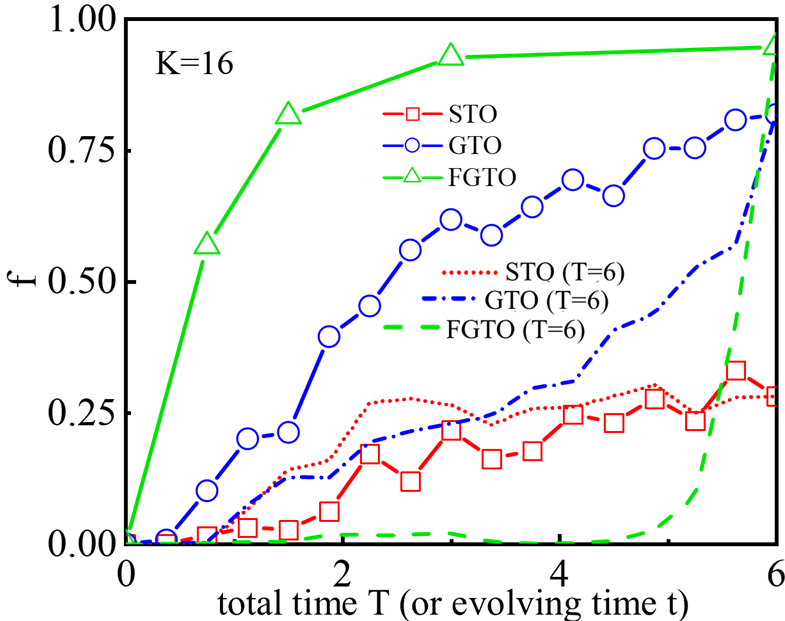}
	\caption{(Color online) The lines with symbols show the \R{fidelity $f$ in Eq. (\ref{eq-f})} vs the total evolution time $T$ for the STO, GTO, and FGTO methods. The target state is the ground state of the Heisenberg model and the evolution Hamiltonian is taken as the XY model. \R{The lines without symbols show $f(t)$ given by Eq. (\ref{eq-Ft}) with $t$ varying from $t=0$ to $t=6$. We fix the number of spins $N=10$ and} the total number of time slices $K=16$ as an example.}
	\label{fig-FvsT}
\end{figure}

In Fig. \ref{fig-FvsT}, \R{we show $f$ versus the total evolution time $T$ by fixing $K=16$ using STO, GTO, and FGTO. By taking a small $T$, say $T\simeq 0.75$ (still with fixed $K=16$ still), the fidelity increases to $f \simeq 0.567$ using FGTO. By increasing $T$, meaning the system will be evolved for a longer time in each time slice, $f$ further increases to $f\simeq 0.217$, $0.618$, and $0.926$ using STO, GTO, and FGTO, respectively, for $T=6$. FGTO obtains the highest fidelity among these three methods.} In other words, FGTO permits the evolution to the target state in the shortest time, in order to reach the same fidelity.

\R{The lines without symbols} illustrates how the state is evolved by showing the \R{fidelity $f$} between the target state and the evolved state at different time $t$, which is defined as
\begin{eqnarray}
	\R{f(t) =\left|\langle \psi_{\text{tar}} | \prod_{k=t/\tau}^{1} e^{-i\tau \hat{H} (k\tau)} |0 \rangle\right|.}
	\label{eq-Ft}
\end{eqnarray}
The magnetic fields in the evolution Hamiltonian are taken as those optimized \R{by the STO, GTO, or FGTO with $T=6$. It is an interesting observation that by FGTO,  $f(t)$ does not monotonously increase with $t$. A slight and smooth decrease of $f(t)$ appears for about $t = 3$. In contrast, the $f(t)$ obtained by GTO or STO in general increases monotonously with $t$. Meanwhile, $f(t)$ achieved by FGTO starts to increase rapidly for $t>5$ approximately. For GTO and STO in comparison, $f(t)$ starts to increase for about $t>1$. For FGTO, the non monotony and long-time evolution before a rapid increase of $f(t)$ might indicate the escape from the local minimum, which results in a higher $f$ compared with that obtained by the STO and GTO algorithms.}

\begin{figure}[tbp]
	\centering
	\includegraphics[angle=0,width=0.8\linewidth]{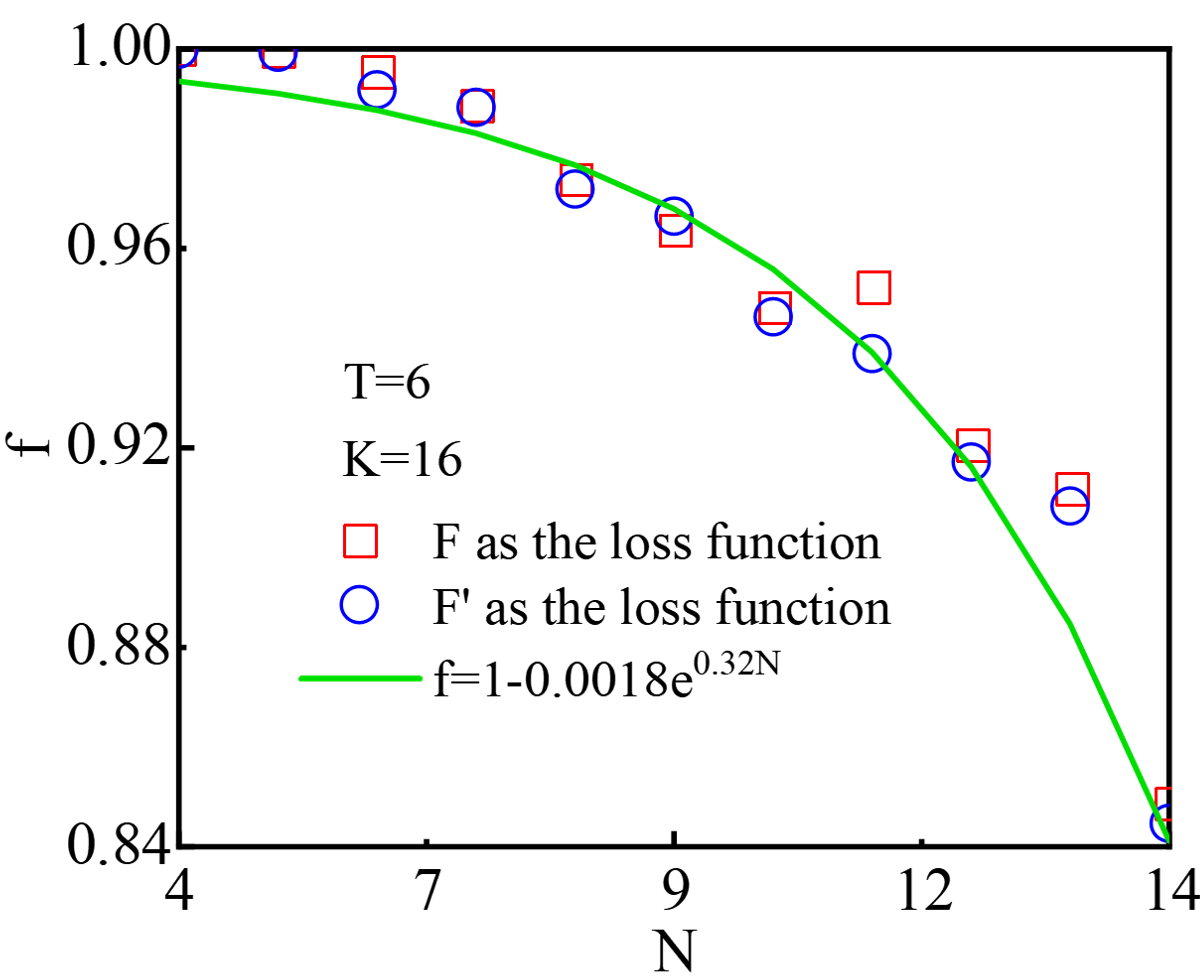}
	\caption{\R{(Color online) The fidelity $f$ in Eq. (\ref{eq-f}) optimized by FGTO vs the number of spins $N$. The data shown by the red squares are obtained using $F$ [Eq. (\ref{eq-F})] as the loss function, and those by the blue circles use $F'$ [Eq. (\ref{eq-1-f})] as the loss. Our results imply the exponential decrease of $f$ against $N$ as $f=1-0.0018e^{0.32N}$. The target state is the ground state of the Heisenberg chain, and the evolution Hamiltonian possesses the XY interactions. We take the total time $T=6$ and $K=16$ time slices for evolution.}}
	\label{fig-fN}
\end{figure}

\R{Figure \ref{fig-fN} shows how the fidelity $f$ between the target and evolved states changes with the number of spins $N$. We use FGTO to optimize the magnetic fields, and fix the total time $T=6$ and the number of time slices $K=16$. We also try to use another loss that is defined as}
\begin{eqnarray}
	\R{F' = 1 - f.}
	\label{eq-1-f}
\end{eqnarray}
\R{The resulting fidelities $f$ are shown by the blue circles in Fig. \ref{fig-fN}. These two loss functions $F$ and $F'$ show similar performance. As expected, $f$ decreases as $N$ increases. Our results suggest an exponential scaling relation between $f$ and $N$ as}
\begin{eqnarray}
	\R{f \simeq 1 - \mu e^{\nu N}.}
	\label{eq-fN}
\end{eqnarray}
\R{In our example where the target is the ground state of the Heisenberg chain and the evolution Hamiltonian possesses the XY interactions, we have $\mu\simeq 1.8 \times 10^{-3}$ and $\nu \simeq 0.32$. These values of $\mu$ and $\nu$ suggest that accurate preparations can be done at least for $N \sim O(1)$. As to the exponential scaling behavior, it might be due to the fact that the fidelity between two states in general decreases exponentially with the number of spins. However, the universality of Eq.( {\ref{eq-fN}}) is to be investigated further in other systems.}

\begin{figure}[tbp]
	\centering
	\includegraphics[angle=0,width=0.8\linewidth]{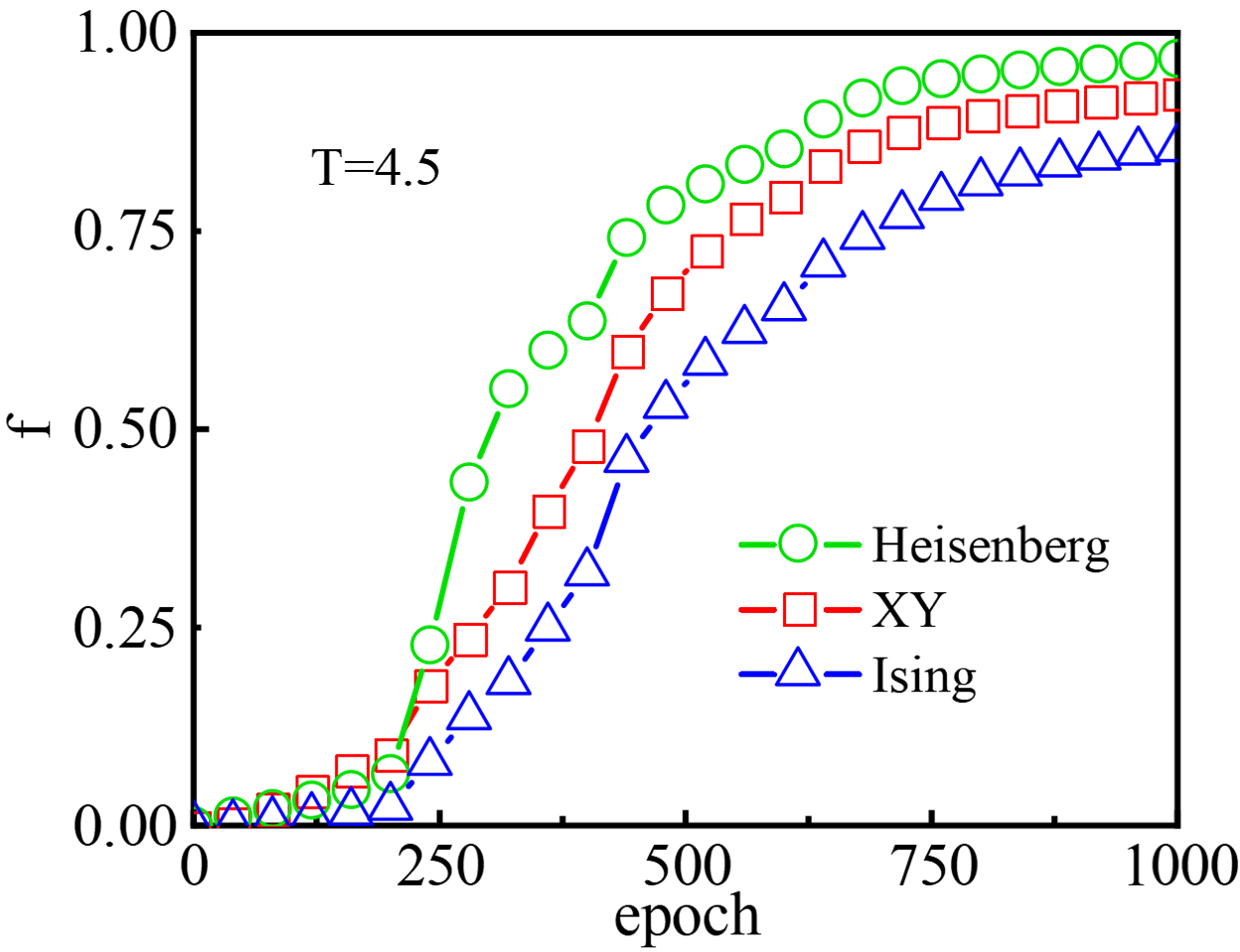}
	\caption{(Color online) The \R{fidelity $f$ in Eq. (\ref{eq-f})} vs the optimization epoch. We take the target state as the ground state of the Heisenberg model, and test the Ising, XY, and Heisenberg models as the evolution Hamiltonians. We fix the total evolution time $T=4.5$ and increase the number of time slices from $K=1$ to $16$ as the optimization proceeds. \R{We take the number of spins as $N=10$}. See details in the main text.}
	\label{fig-threeModel}
\end{figure}

To demonstrate how the evolved state approaches the target state with the optimization in FGTO, Fig. \ref{fig-threeModel} gives the \R{fidelity $f$} at different optimization epochs. Specifically, we start with a $K=1$ time slice.  The magnetic fields are optimized for $200$ epochs and afterward we fine grain the time discretization by increasing $K$ to $2K$. In total,  we implement 1000 epochs and eventually have $K=16$ time slices. The total evolution time is fixed to be $T=4.5$.

Our results show that not just the XY model but also the Ising model with only the interactions along the spin-z direction can be used as the evolution Hamiltonian to prepare the ground state of the Heisenberg model. We speculate that any Hamiltonian that can entangle the whole system by time evolution could be used as the evolution Hamiltonian for state preparation. As the magnetic fields correspond to single-body operators and cannot produce any entanglement, the coupling terms (two-body in our examples) with fixed strength are to entangle the system, and the magnetic fields drive the entangled state to the target. By introducing more interaction terms in the evolution Hamiltonian, such as the Heisenberg model that contains the interactions along all three spin directions, one could obtain higher precision with the same total evolution time and number of time slices. We expect higher precision by using the evolution Hamiltonian that can entangle the system in a higher speed, such as those with proper longer-range or multi-body interactions.

\begin{figure}[tbp]
	\centering
	\includegraphics[angle=0,width=1\linewidth]{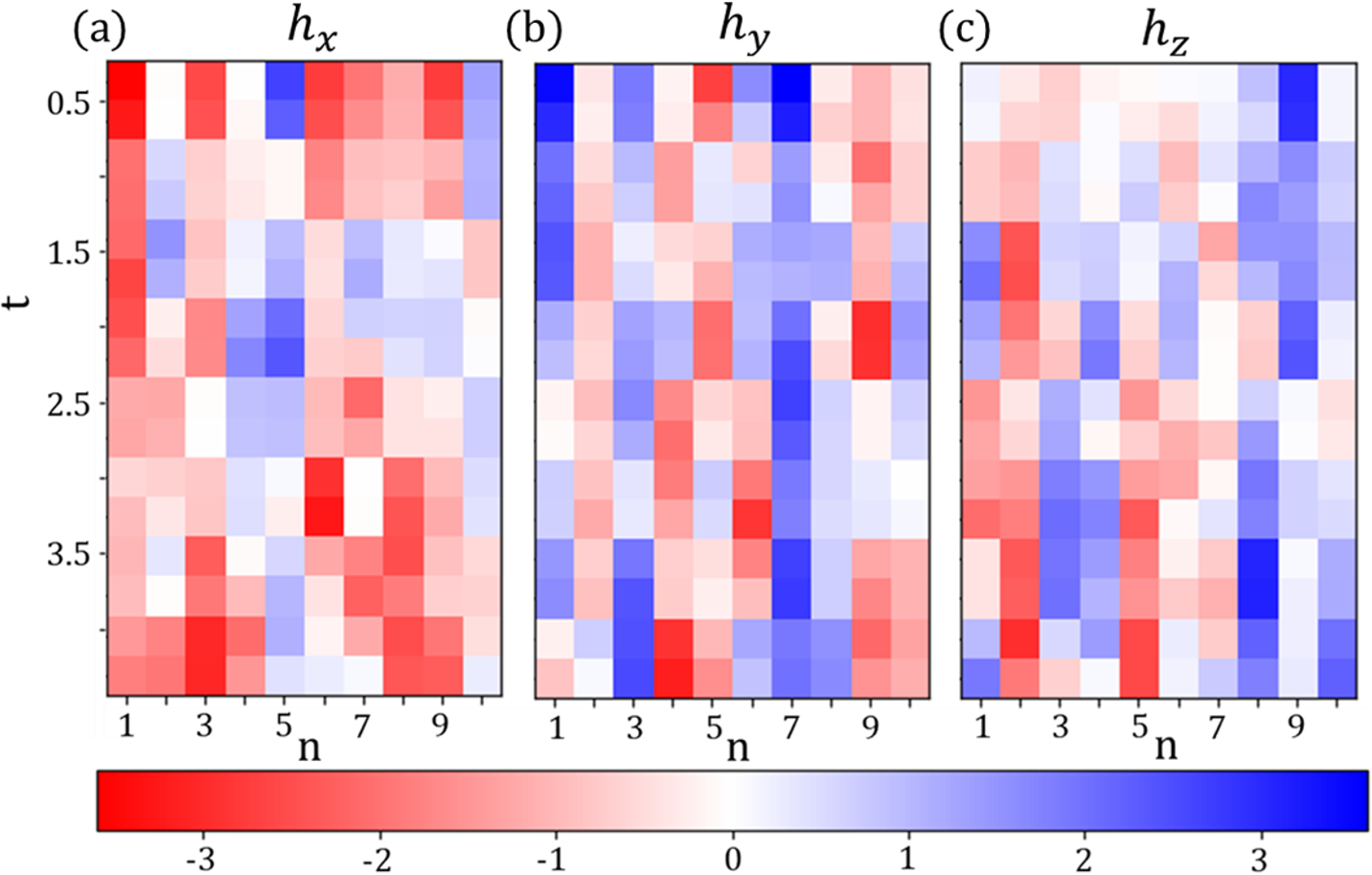}
	\caption{(Color online) The space-and-time ``landscapes'' of the magnetic fields in the (a) x, (b) y, and (c) z directions for preparing the ground state of the Heisenberg chain by time evolution with XY interactions. We have $t=k\tau$ that denotes the time in the evolution and use $n$ to denote the position in the chain. The colors illustrate the strength of the magnetic fields.} See details in the main text.
	\label{fig-h}
\end{figure}

Figure \ref{fig-h} demonstrates the space-and-time landscapes of the magnetic fields in three directions, obtained by the FGTO methods. We take the XY model as the evolution Hamiltonian and the ground state of the Heisenberg model as the target. The color of the block at the $n$-th column and the $k$-th row illustrates the strength of the magnetic field on the $n$-th spin within the $k$-th time slice. We take the total evolution time as $T=4.5$ and the total number of time slices $K=16$. The initial product state will be evolved to the target by imposing the magnetic fields according to this figure.

\begin{figure}[tbp]
	\centering
	\includegraphics[angle=0,width=0.8\linewidth]{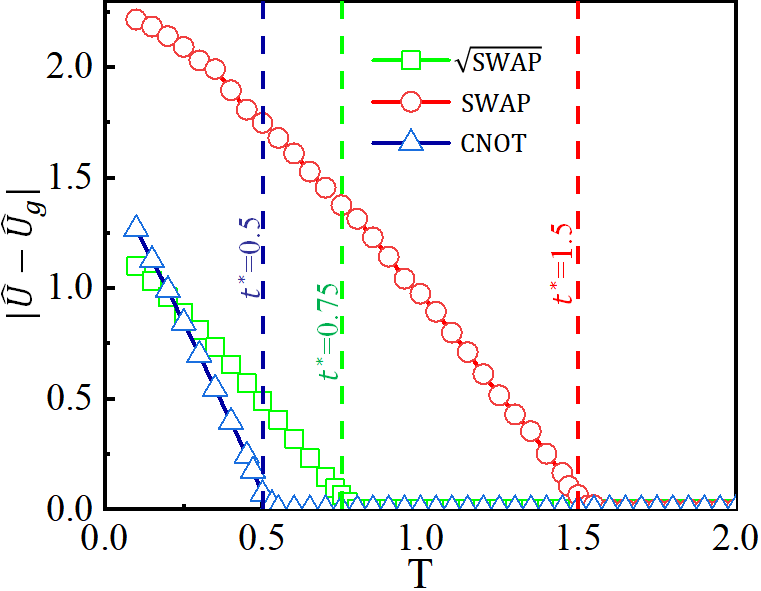}
	\caption{ The difference between the time-evolution gate and the target gate with different evolution times $T$. The SWAP, $\sqrt{\text{SWAP}}$, and CNOT gates are taken as the target as examples. The minimal time $t^{\ast}$ given by the analytical solution for realizing each gate is indicated by the vertical dash lines.}
	\label{fig-gates}
\end{figure}

To further test our method, we optimize the magnetic fields to realize a target gate $\hat{G}_{\text{tar}}$. The loss function is defined as the difference better the time-evolution gate $\hat{G}(T) $ and the target as
\begin{eqnarray}
	F_{G} = \left|\hat{G}(T) - \hat{G}_{\text{tar}} \right|.
	\label{eq-gateloss}
\end{eqnarray}
We choose $\hat{G}_{\text{tar}}$ as a two-qubit gate (SWAP, $\sqrt{\text{SWAP}}$, or CNOT) as the minimal time $t^{\ast}$ to obtain it by quantum control can be analytically given\cite{2001Time, 2013Time, 2020Time}. GTO is used considering we do not encounter the problems of local minima in these cases. As shown in Fig. \ref{fig-gates}, the difference becomes close to zero at the minimal time given by the analytical solution.  The converged difference scales as the learning rate $\sim O(\eta)$ due to the optimization fluctuations. These results shows that the optimal control are reached by our scheme.

\section{Discussions and perspective}

In this work, we consider the preparation of the ground states of quantum many-body systems by time evolution with fixed spin couplings and tunable magnetic fields. We focus on the cases where the Hamiltonian for the controlled time evolution contains less interaction terms than the Hamiltonian that gives the ground state as the target. The fine-grained time optimization (FGTO) algorithm is proposed to gain high efficiency and stability. We test our proposal by preparing the ground state of the Heisenberg model with the time evolution by only the XY or Ising interactions. FGTO achieves high fidelity compared with two baseline methods using local and global optimization strategies, respectively. Our work can be readily generalized to the preparation of not just many-body ground states, but also the states constructed by hand for the purposes of realizing nontrivial physical properties or implementing the tasks of quantum communication and computation.

From a theoretical perspective, it is interesting and important to study the completeness. Usually, it is difficult to adjust or control the interaction terms in quantum many-body systems. \R{There exist considerable experiments on the lattice models of spin-1/2's on the artificial platforms by quantum simulation, such as cold atoms~\cite{ExpSpin1, ExpSpin2, ExpSpin3} and superconducting circuits~\cite{ExpSC1, ExpSC2, ExpSC3, ExpSC4}. However, } the interactions appearing in the materials or simulators at hand might suffer strict restrictions on, e.g., the interaction form or range. Our work implies certain generality of using restricted interactions for state preparation, or helps to reduce the complexity of the evolution Hamiltonian. It is an open and important issue to investigate and characterize the set of classes reachable by time evolution with the given interactions. \R{In our cases, the Hamiltonians for evolution are imposed with several restrictions. Only certain fixed two-body terms (interactions) and variational coefficients of the single-body terms (magnetic fields) are contained in the Hamiltonians. All operators are Pauli operators, i.e., the SU(2) group, which is a very small part among the generators of all possible unitary operators in the Hilbert space. The two-body terms only allow nearest-neighbor interactions in a 1D chain. It is interesting to investigate how different restrictions would affect the range of accessible target states and possibly the scaling between $f$ and $N$ [as Eq. (\ref{eq-fN})].}

By modifying the loss function, our work can be used to prepare the states the possess certain desired physical properties such as those with high entanglement~\cite{QCent2010PRL, QCent15PRA, QCent2015PRA2, QCentPRA2017}. Our idea can also be generalized with non-Hermitian time evolution~\cite{nonHermitian2007}, and applied to study the inhomogeneous Kibble-Zurek mechanism in many-body systems, including the annealing processes and phase transitions affected by spatial inhomogeneity~\cite{KZM2013JPCM, KZM2016JPCM5, KZM2018JPCM2, KZM2018JPCM3, KZM2018JPCM4, KZM2019JPCM1}. \R{More properties of the preparation scheme, such as sensitivity to magnetic fields at different times, can be probed by employing the Sobol-indices-based methods ~\cite{im1993sensitivity, e72a2462c52d404d88e48b5eecc5fc34}. It is also an interesting topic to probe the time correlations with the expansion of the high-dimensional model representation ~\cite{2001High}.}

\section*{Acknowledgment} The authors are thankful to Heng Fan, Gang Su, Shao-Ming Fei, Zhi-Yuan Ge, and Yang Zhao for stimulating discussions. This work was supported by NSFC (Grants No. 12004266 and No. 11834014), Beijing Natural Science Foundation (No. 1192005 and No. Z180013), Foundation of Beijing Education Committees (No. KM202010028013), and the key research project of Academy for Multidisciplinary Studies, Capital Normal University.

\bibliography{bibi}
\end{document}